\documentclass{aastex61}

\usepackage[T1]{fontenc}
\usepackage{ae,aecompl}
\usepackage{footnote}
\usepackage{ctable}
\usepackage{graphicx}	
\usepackage{amsmath}	
\usepackage{amssymb}	

\usepackage[normalem]{ulem}     

\newcommand\aastex{AAS\TeX}

\newcommand{\angstrom}{\mbox{\normalfont\AA}}

\received{December, 2017}
\revised{xxxx}
\accepted{\today}
\submitjournal{ApJ}

\shorttitle{\aastex\ Abundances in EFRs}
\shortauthors{Baker et al.}

\begin{document}

\title{Coronal Elemental Abundances in Solar Emerging Flux Regions}

\correspondingauthor{Deborah Baker}
\email{deborah.baker@ucl.ac.uk}

\author[0000-0002-0665-2355]{Deborah Baker}
\affil{University College London, Mullard Space Science Laboratory, Holmbury St. Mary, Dorking, Surrey, RH5 6NT, UK}

\author[0000-0002-2189-9313]{David H. Brooks}
\affiliation{College of Science, George Mason University, 4400 University Drive, Fairfax, VA 22030, USA}

\author[0000-0002-2943-5978]{Lidia van Driel-Gesztelyi}
\affiliation{University College London, Mullard Space Science Laboratory, Holmbury St. Mary, Dorking, Surrey, RH5 6NT, UK}
\affiliation{Observatoire de Paris, LESIA, UMR 8109 (CNRS), F-92195 Meudon, France}
\affiliation{Konkoly Observatory of the Hungarian Academy of Sciences, Budapest, Hungary}

\author[0000-0001-7927-9291]{Alexander W. James}
\altaffiliation{University College London, Mullard Space Science Laboratory, Holmbury St. Mary, Dorking, Surrey, RH5 6NT, UK}

\author[0000-0001-8215-6532]{Pascal D\'emoulin}
\affiliation{Observatoire de Paris, LESIA, UMR 8109 (CNRS), F-92195 Meudon, France}

\author[0000-0003-3137-0277]{David M. Long}
\affiliation{University College London, Mullard Space Science Laboratory, Holmbury St. Mary, Dorking, Surrey, RH5 6NT, UK}

\author{Harry P. Warren}
\affiliation{Space Science Division, Naval Research Laboratory, Washington DC 20375, USA}

\author{David R. Williams}
\affiliation{European Space Agency / ESAC, E-28692 Villanueva de la Ca\~nada, Madrid, Spain}

\begin{abstract}

The chemical composition of solar and stellar atmospheres differs from that of their photospheres. 
Abundances of elements with low first ionization potential (FIP) are enhanced in the corona relative to high FIP elements with respect to the photosphere. 
This is known as the FIP effect and it is important for understanding the flow of mass and energy through solar and stellar atmospheres. 
We used spectroscopic observations from the Extreme-ultraviolet Imaging Spectrometer (EIS) onboard the \emph{Hinode} observatory to investigate the spatial distribution and temporal evolution of coronal plasma composition within solar emerging flux regions inside a coronal hole. 
Plasma evolved to values exceeding those of the quiet Sun corona during the emergence/early decay phase at a similar rate for two orders of magnitude in magnetic flux, a rate comparable to that observed in large active regions containing an order of magnitude more flux. 
During the late decay phase, the rate of change was significantly faster than what is observed in large, decaying active regions. 
Our results suggest that the rate of increase during the emergence/early decay phase is linked to the fractionation mechanism leading to the FIP effect, whereas the rate of decrease during the later decay phase depends on the rate of reconnection with the surrounding magnetic field and its plasma composition.

\end{abstract}

\keywords{Sun: abundances -- Sun: corona -- Sun: evolution -- Sun: magnetic fields}

\section{Introduction} \label{sec:intro}
Element abundance patterns have long been used as tracers of physical processes throughout astrophysics.  
The benchmark reference for all cosmic applications is the solar chemical composition.
The observed variation in coronal solar and stellar abundances depends on the first ionization potential (FIP) of the main elements found in the solar atmosphere.
Those elements with FIP greater than 10 eV (high-FIP elements) maintain their photospheric abundances in the corona whereas elements with lower FIP have enhanced abundances (low-FIP elements), \emph{i.e.,} the so-called solar/stellar FIP effect.
Conversely, the inverse FIP (IFIP) effect refers to the enhancement/depletion of high-/low-FIP elements in solar and stellar coronae. 
FIP bias is the factor by which low-FIP elements such as Si, Mg, and Fe are enhanced or depleted in the corona relative to their photospheric abundances.

\cite{wood10} carried out a survey of FIP bias in quiescent stars with X-ray luminosities less than 10$^{29}$ erg s$^{-1}$.
The selection criteria excluded the most active stars.
They found a clear dependence of FIP bias on spectral type in their sample of 17 G0--M5 stars.
As the spectral type becomes later (G$\rightarrow$K$\rightarrow$M), the FIP effect observed in G-type stars, including the Sun, decreases to zero (at about K5) then becomes the IFIP effect for M dwarfs.
\cite{laming15}, and references therein, updated and extended the sample of \cite{wood10}, finding that the trend remains the same for less active stars.
When the Sun is observed as a star, \emph{i.e.,} observed as an unresolved point source, FIP bias is $\sim$3--4 \citep[for a single measurement made during solar maximum; ][]{laming95}, similar to solar analogs $\chi$$^{1}$ Ori \citep[$\sim$3;][]{telleschi05} and $\alpha$ Cen A \citep[$\sim$4;][]{raassen03}. 
Using a long-term data series of Sun-as-a-star observations, \cite{brooks17} demonstrate a high correlation between the variations of coronal composition and the phase of the solar cycle using the 10.7 cm radio flux as solar activity proxy.
Recently, \cite{doschek15} and \cite{doschek16,doschek17} used spatially resolved spectroscopic observations to provide the first evidence of the IFIP effect on the Sun in the flare spectra of flaring active regions near large sunspots.
The values of solar IFIP in these specific locations are similar to the levels of IFIP observed in the M dwarf stars of \cite{wood10} and \cite{laming15}.

Understanding the spatial and temporal variation of elemental abundances in the solar corona provides insight into how matter and energy flow from the chromosphere into the heliosphere.
The fractionation of plasma takes place in the chromosphere where high-FIP elements are mainly neutral and low-FIP elements are ionized mostly to their 1$^{+}$ or 2$^{+}$ stages.
In fact, the fractionation process is likely to be related to the production of coronal plasma \citep{sheeley96}.
It is thought that the enhancement of low-FIP elements arises from the ponderomotive force due to Alfv\'enic waves acting on chromospheric ions that are preferentially accelerated into the corona while neutral elements remain behind \citep{laming04,laming09,laming12,laming15,dalhburg16}.
Ponderomotive acceleration occurs close to loop footpoints, more precisely in the upper chromosphere, where the density gradient is steep.
For the first time, \cite{dalhburg16} have demonstrated that ponderomotive acceleration occurs at the footpoints of  coronal loops as a by-product of coronal heating in their 3D compressible MHD numerical simulations.

Solar plasma composition may be used as a tracer of the magnetic topology in the corona \citep{sheeley95} and as a means to link the solar wind to its source regions \citep{gloeckler89,fu17}.
The magnetic field of coronal holes (CHs) extends into the solar wind and is deemed to be open field. 
Such fields are observed to contain unenriched photospheric plasma \citep{feldman98,brooks11,baker13}, as does the fast solar wind emanating from CHs \citep{gloeckler89}. 
The closed field in the cores of quiescent active regions (ARs) holds plasma with FIP bias of about 3, \emph{i.e.,} coronal composition \citep{delzanna14}.
Quiet Sun field also has enriched plasma with FIP bias $\sim$1.5--2 \citep[\emph{e.g., }][]{warren99,baker13,ko16}.
Blue-shifted upflows located at the periphery of ARs have enhanced FIP bias of 3--5 \citep[\emph{e.g., }][]{brooks11,brooks12a}.
These upflows may become outflows and make up a part of the slow solar wind \citep{brooks11,lvdg12,culhane14,mandrini14,brooks15,fazakerley16}.

Until recently, knowledge of FIP bias evolution was largely based on spectroheliogram observations from \emph{Skylab} in the 1970s.
In these observations, newly emerged flux is photospheric in composition \citep{sheeley95,sheeley96,widing97,widing01}.
The plasma then progressively evolves to coronal FIP bias levels of $\sim$3 after two days and exceeding 7--9 after 3--7 days \citep{feldman00,widing01}.
Recent spectroscopic observations have revealed a different and more complex scenario for FIP bias evolution during the later stages of AR lifetimes.
\cite{baker15} found that FIP bias in the corona decreases in a large decaying AR over 2 days as a result of the small-scale evolution in the photospheric magnetic field.
Flux emergence episodes within supergranular cells modulate the AR's overall plasma composition.
Only areas within the AR's high flux-density core maintain coronal FIP bias.
In another large decaying AR, \cite{ko16} also observe a decline in FIP bias over five days before reaching a basal state in the quiet Sun (FIP bias $\sim$1.5).  
This occurred as the photospheric magnetic field weakened over the same time period.

In this work, we exploit a series of spectroscopic observations to study the time evolution and distribution of plasma composition in emerging flux regions (EFRs) of varying magnetic fluxes from ephemeral regions to pores without spots and an active region with spots.
Throughout the paper, we use the term EFRs when referring to the coronal counterpart of magnetic bipoles observed in the photosphere. 
These regions emerge, evolve, and decay within the open magnetic field of a CH located at low solar latitude. 
Moreover, we use `enriched plasma' when referring to plasma composition of FIP bias 2--3$^{+}$ \emph{i.e.} well in excess of quiet Sun FIP bias ($\sim$1.5).
Low-FIP elements are enhanced compared with high-FIP elements. 
We provide a brief description of the coronal extreme-ultraviolet (EUV) and magnetic field observations in Section \ref{obs}.
This is followed by a full account of the method of analysis in Section \ref{method}.
Results are presented in Section \ref{results}.
We discuss our results, especially within the context of the original \emph{Skylab} observations, in Section \ref{disc}.
Finally, we summarize our results and draw our conclusions in Section \ref{conclusion}.

\section{Observations}\label{obs}
Two solar satellite observatories, the \emph{Solar Dynamics Observatory} \citep[SDO; ][]{pesnell12} and \emph{Hinode} \citep{kosugi07}, provided the observations used in this study.
SDO's Atmospheric Imaging Assembly \citep[AIA; ][]{lemen12} images the solar atmosphere in ten passbands in the  temperature range from 5 $\times$ 10$^{3}$ K to 2 $\times$  10$^{7}$ K. 
The Helioseismic and Magnetic Imager \citep[HMI; ][]{schou12,scherrer12} onboard SDO makes high-resolution measurements of the line of sight and vector magnetic field of the solar surface.
Both SDO instruments observe the full solar disk at unprecedented temporal and spatial resolutions.
The Extreme-ultraviolet Imaging Spectrometer \citep[EIS; ][]{culhane07} onboard \emph{Hinode} is a normal incidence spectrometer that obtains spatially resolved spectra in two wavelength bands: 170--210 $\angstrom$ and 250--290 $\angstrom$ with a spectral resolution of 22 m$\angstrom$.
Spectral lines are emitted in the EUV at temperatures ranging from 5 $\times$ 10$^{4}$ K to 2 $\times$ 10$^{7}$ K.
The limited field of view of EIS is constructed by rastering the 1$\arcsec$ or 2$\arcsec$ slit in the solar West-East direction.

\subsection{Coronal EUV Observations} 
\begin{figure}
	\includegraphics[width=\columnwidth]{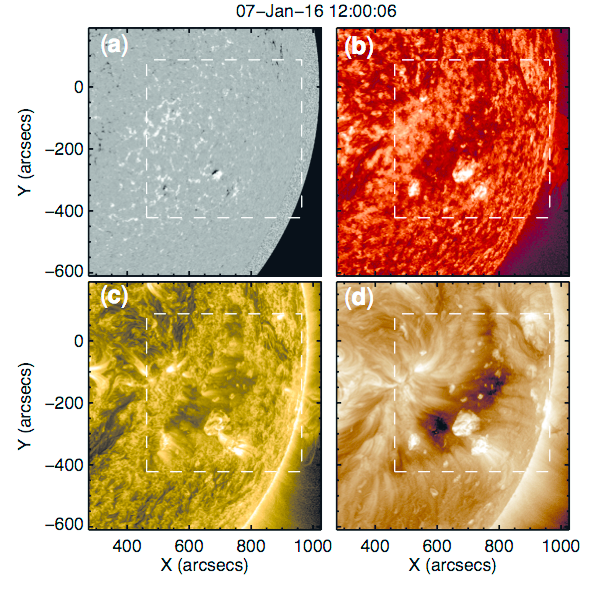}
    \caption{SDO/HMI line-of-sight magnetogram (a) and AIA 304~\angstrom~(b), 171~\angstrom~(c), and 193~\angstrom~(d) high resolution images at 12:00 UT on 7 January 2016.  \emph{Hinode}/EIS field of view is indicated by the white box.  Coordinates are in arcsec  ($\arcsec$) with origin set at the solar disk center.  High resolution images are made using the Multi-Gaussian Normalization (MGN) technique \citep{morgan14}.  (Figure is from Movie 1).  }
    \label{fig:aia}
\end{figure}

A small CH, located in the southern hemisphere (SW quadrant), was observed by SDO/AIA and SDO/HMI beginning on 6 November 2015.  
For the next two solar rotations, the CH evolved and approximately doubled in area.  
By the time it crossed the solar central meridian on 3 January 2016, it had 
extended by $\sim$200$\arcsec$ into the northern hemisphere to form a narrow channel connecting to the northern polar CH.  
Figure \ref{fig:aia} contains a 4-panel (a--d) image of the CH from supplementary material Movie 1:  (a) SDO/HMI magnetogram, (b) SDO/AIA 304 $\angstrom$, (c) 171 $\angstrom$, and (d) 193 $\angstrom$ images with the EIS field of view indicated in each panel by the white dashed box. 
The movie covers the period from 16:00 UT on 4 January 2016 to 23:30 on 7 January 2016.  
Throughout this period, the environment of the CH was highly dynamic with repeated flux emergence, flux cancellation, jets, brightenings, and flaring.

\subsection{EUV Spectroscopic Observations} 
\begin{figure*}
	\centering
    \includegraphics[width=6.5in,trim = {0.2in 2.2in 0.2in 2.3in }, clip]{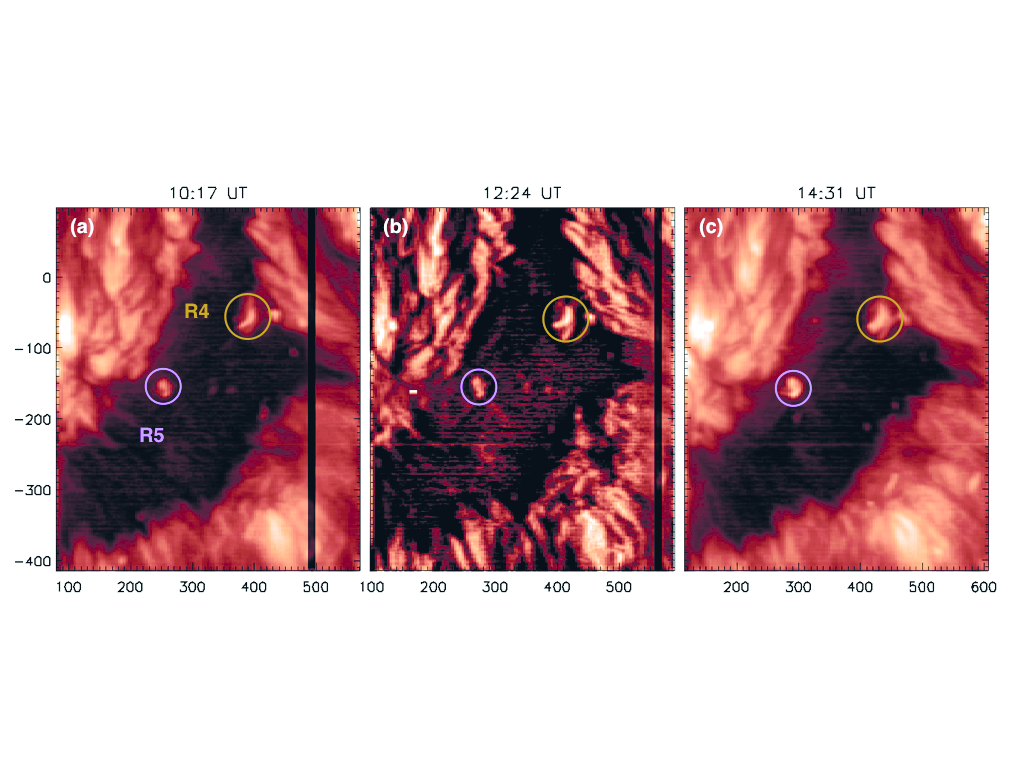}
    \includegraphics[width=6.5in,trim = {0.2in 2.2in 0.2in 2.3in }, clip]{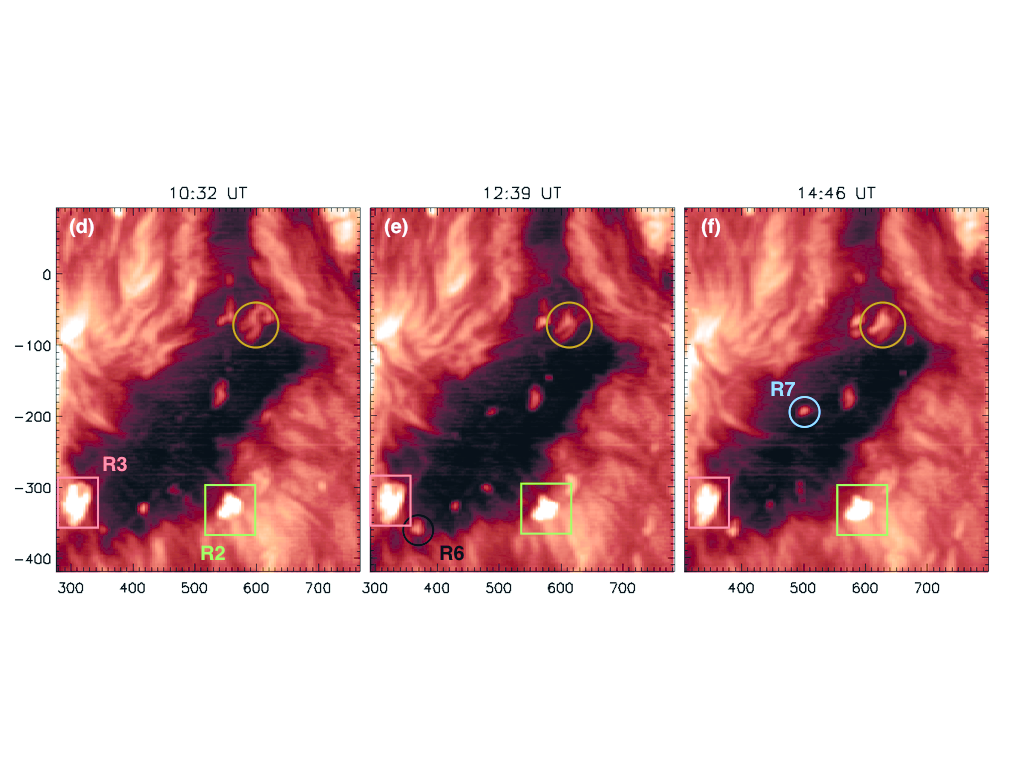}
    \includegraphics[width=6.5in,trim = {0.2in 2.2in 0.2in 2.3in }, clip]{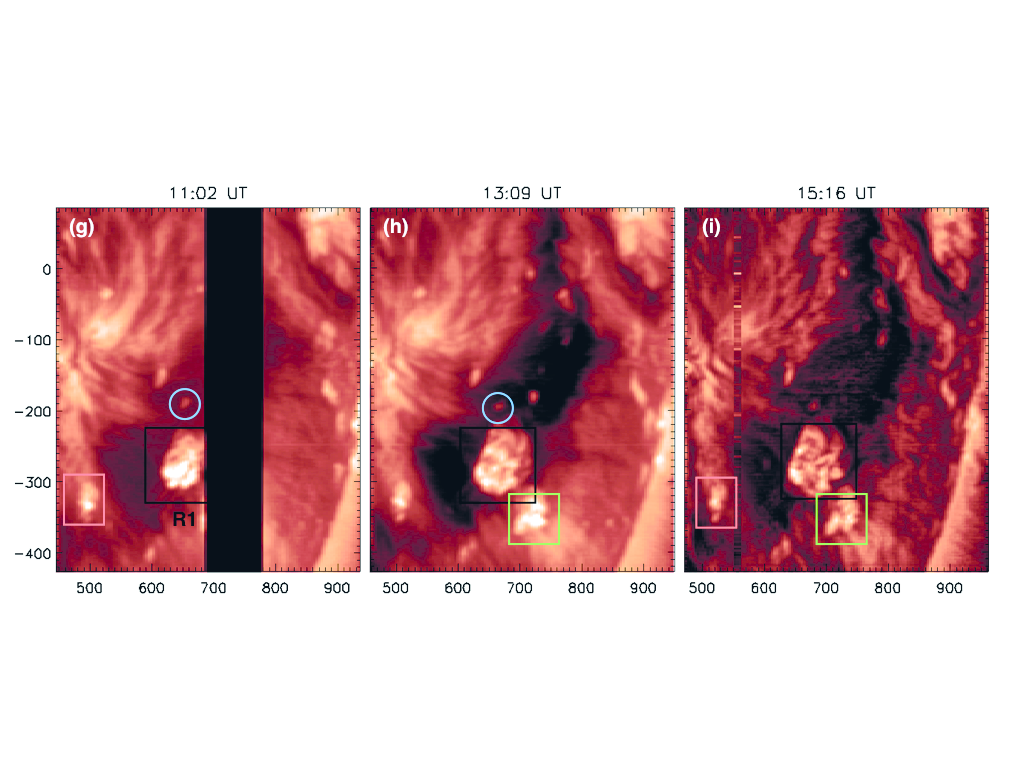}
    \caption{ \emph{Hinode}/EIS Fe {\sc xii} 195.12~\angstrom~intensity maps for 5 (top), 6 (middle), and 7 (bottom) January 2016.  Regions R1--R7 in the text and Table \ref{tab:results} are encircled/surrounded in the EIS maps for which FIP measurements are made (colors correspond to Figure \ref{fig:plots}:  R1 - black box, R2 - green box, R3 - red box , R4 - tan circle, R5 - purple circle, R6 - black circle, R7 blue circle).  Coordinates are in arcsec ($\arcsec$)  with origin at solar disk center.  Images are enhanced using the Multi-scale Gaussian Normalization technique \citep[MGN; ][]{morgan14}.  Figure \ref{fig:mini_fip2} contains zoomed intensity maps which correspond to R3 in panel (d) and R1--R2 in panel (h).} 
    \label{fig:obs}
\end{figure*}

\emph{Hinode}/EIS tracked the CH and its surroundings from 5--7 January 2016 taking three successive large field-of-view rasters during the South Atlantic Anomaly (SAA)-free periods on each day.
Figure \ref{fig:obs} displays the Fe {\sc xii} 195.12 $\angstrom$ high resolution intensity images from each raster enhanced using the Multi-scale Gaussian Normalization technique of \cite{morgan14}.
The field of view measuring 492$\arcsec$ $\times$ 512$\arcsec$ was constructed by rastering the 2$\arcsec$ slit in 4$\arcsec$ coarse steps.
At each pointing position, EIS took 60 s exposures so that the total raster time exceeded two hours.
The study used for these observations was specially designed for abundance measurements \citep{brooks15}.
It contains an emission line from both a high FIP element (S {\sc x} 264.233 $\angstrom$) and a low FIP element (Si {\sc x} 258.375 $\angstrom$) formed at approximately the same temperature (1.5 $\times$ 10$^{6}$ K).
To account for residual temperature and density effects on the line ratio, the study includes a series of Fe lines ({\sc viii - xvi}) covering 0.45 $\times$ 10$^{6}$ K to 2.8 $\times$ 10$^{6}$ K to determine emission measure distributions and density sensitive line-pair Fe {\sc xiii}.
Details of the \emph{Hinode}/EIS study are given in Table \ref{tab:study}. 

\begin{table}
	\centering
	\caption{\emph{Hinode}/EIS Study Details.}
	\label{tab:study}
	\begin{tabular}{ll} 
		\hline
		Study number & 513\\
		Emission Lines & Fe {\sc viii} 185.213 $\angstrom$, Fe {\sc ix} 188.497 $\angstrom$\\
        &Fe {\sc x} 184.536 $\angstrom$, Fe {\sc xi} 188.216 $\angstrom$\\
        &Fe {\sc xi} 188.299 $\angstrom$, Fe {\sc xii} 192.394 $\angstrom$\\
        &Fe {\sc xii} 195.119 $\angstrom$, Fe {\sc xiii} 202.044 $\angstrom$\\
        &Fe {\sc xiii} 203.826 $\angstrom$, Fe {\sc xiv} 264.787 $\angstrom$\\
        &Fe {\sc xv} 284.16 $\angstrom$,Fe {\sc xvi} 262.984 $\angstrom$\\
        &Si {\sc x} 258.375 $\angstrom$, S {\sc x} 264.233 $\angstrom$\\
		Field of view & 492$\arcsec$ $\times$ 512$\arcsec$\\
		Rastering & 2$\arcsec$ slit, 123 positions, 4$\arcsec$ coarse steps \\
        Exposure Time &60 s\\
        Total Raster Time &2 h 7 m\\
		\hline
	\end{tabular}
\end{table}


\subsection{Magnetic Field Observations}
Figure \ref{fig:aia} (a) shows the line-of-sight magnetic field component in the SDO/HMI magnetogram on 7 January at 12:00 UT.
The CH field was dominantly comprised of positive polarity magnetic field.
The average magnetic flux density was $\sim$11 G, which is typical of CHs for the declining phase of the solar cycle during which the measurements were made \citep{harvey82,hofmeister17}.
A number of small bipoles emerged and/or decayed into the positive polarity CH throughout the observing period.
The range of peak flux of the bipoles extended to more than three orders of magnitude from $10^{18}$ Mx to greater than $10^{21}$ Mx.
The bipoles evolved in different magnetic topological environments.
Some emerged fully within the CH so the topological environment was open whereas other bipoles were located along the CH boundary so that one polarity was in close proximity to open magnetic field and the opposite polarity was nearby closed field.
Three of the largest EFRs are close to the southern edge of the \emph{Hinode}/EIS field of view in Figure \ref{fig:aia}.

\section{Method of Analysis}\label{method}

\subsection{Region Definition}\label{region_define}
We identified seven EFRs in the CH and along its boundary with the quiet Sun.
In Figure \ref{fig:obs}, they are labelled R1 -- R7 in descending order of maximum magnetic flux.
Defining and tracking these regions in the \emph{Hinode}/EIS observations required great care as their average FIP bias values were sensitive to the extent of, and features contained within, the selected areas. 
To minimize the effect of non-EFR pixels, individual regions were extracted from each raster image using a histogram-based intensity thresholding technique similar to the one employed by \cite{krista09}. 
Intensity thresholding provides a quantitative method for identifying the main features within each of the nine rasters.
We were able to distinguish CH and quiet Sun pixels from EFR pixels within the selected area using this method. 

\begin{figure}
\centering
	\includegraphics[width=0.6\columnwidth]{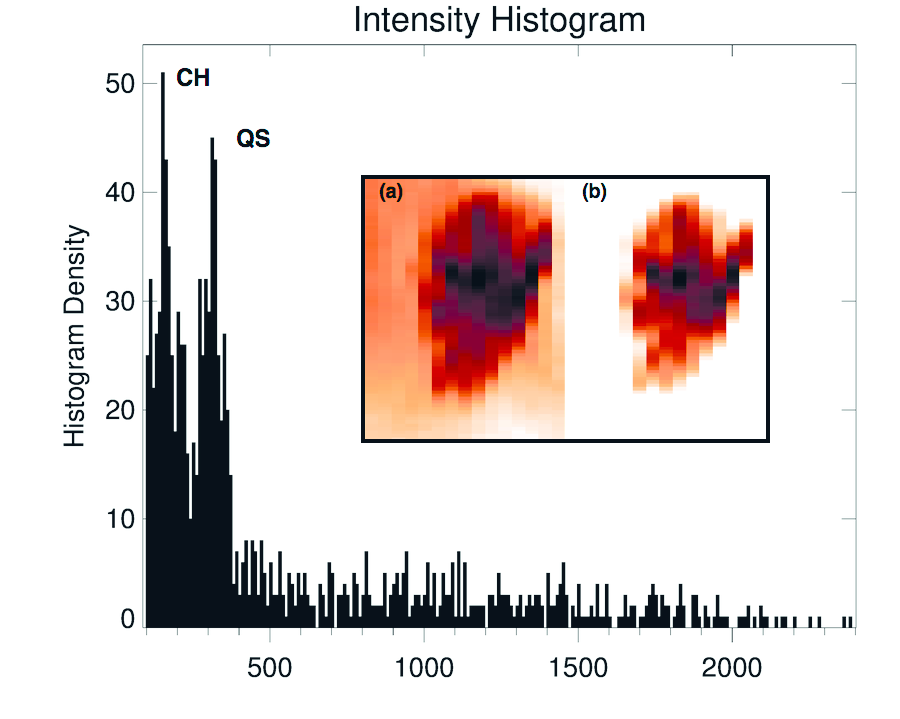}
    \caption{Histogram of Fe {\sc xii} intensity in erg/cm$^{2}$/s/sr within the submap of EFR R3 (left panel of the inset).  Coronal hole (CH) and quiet Sun (QS) intensity peaks are indicated.  The local minimum after the quiet Sun peak at 400 is the intensity threshold, below which, pixels are masked from the region.  FIP measurements are determined by averaging spectra of the unmasked pixels with intensity greater than 400 (right panel of the inset). Unmasked (a) and masked (b) intensity maps are displayed in reverse colors so that EFR features are dark and the surrounding CH is light.}
    \label{fig:mask}
\end{figure}

A histogram of EFR R3 Fe {\sc xii} 195.12 $\angstrom$ intensity at 14:46 UT on 6 January is shown in Figure \ref{fig:mask} as an example.
The trimodal distribution corresponds to the CH, quiet Sun, and EFR within the image of the area surrounding R3 (see the reverse color image in the left panel of the inset in Figure \ref{fig:mask}).  
The first two modes of the histogram, with intensity peaks at $\sim$150 and $\sim$320, are associated with the CH and quiet Sun portions of the distribution, respectively. 
The local minimum at $\sim$400 occurring after the second peak was used to establish the cut-off threshold intensity level for EFR R3 in this raster. 
Pixels with values below the cut off were masked from the region.
The unmasked (a) and masked (b) quiet Sun and CH pixels/areas are shown in the inset of Figure \ref{fig:mask}.
EFR R3 was located at the edge of the CH (middle and bottom rows of Figure \ref{fig:obs}), close to the  nearby quiet Sun so a trimodal intensity distribution is expected.
For EFRs located entirely within the CH, the intensity histogram was bimodal, one mode each for the CH  and EFR pixels.
The cut off threshold was then the local minimum after the CH peak in the intensity distribution. 
The intensity thresholds used to identify CH, quiet Sun, and EFR for R1--R7 varied by $\sim$20$\%$, depending on whether the EFR was fully surrounded by CH or a mix of CH and quiet Sun.
A number of EFRs were excluded from the study because they contained too few EFR intensity pixels ($<50$) \emph{e.g.} R7 in Figure \ref{fig:obs}(e) and (i) and R6 in Figure \ref{fig:obs}(f).

The effect of instrumental stray light is estimated to be $\sim$2$\%$ of the typical counts for Fe {\sc xii} observations of the quiet Sun (see \emph{Hinode}/EIS Software Note No. 12).
For CH observations, the effect is likely to be slightly higher.  
Based on the intensity histogram in Figure \ref{fig:mask}, the threshold for CH pixels is 150 erg/cm$^{2}$/s/sr compared to 400 erg/cm$^{2}$/s/sr for quiet Sun pixels, therefore we estimate the stray light effect to be 4--5$\%$.
In the case of the Si {\sc x}--S {\sc x} FIP bias ratio, the effect is present in both lines, therefore we would expect a differential effect which is lower than that for a single ion.
Moreover, the overall effect of stray light on the FIP bias measurements will be negligible since we are removing CH pixels and using only EFR pixels by the masking process.

\subsection{FIP Bias Measurements} \label{fipmeasure}
\emph{Hinode}/EIS data reduction was carried out using standard routines which are available in Solar SoftWare \citep[SSW;][]{freeland98}.  The eis$\_$prep routine converts the CCD signal in each pixel into calibrated intensity units and removes/flags cosmic rays, dark current, dusty, warm and hot pixels.
All data were corrected for instrumental effects as follows.  The orbital spectrum drift and CCD spatial offsets were corrected using the artificial neural network model of \cite{kamio10}. 
The grating tilt was adjusted using the eis$\_$ccd$\_$offset routine. 
Finally, the CCD detector sensitivity was corrected using the method of \cite{delzanna13} which assumes no degradation after September 2012.

All spectral lines from consecutive ionization stages of Fe {\sc viii-xvi} were fitted with single Gaussian functions with the exception of the known blends of the Fe {\sc xi, xii, xiii} lines for which multiple Gaussian functions were necessary to distinguish the blended lines.
Single Gaussian functions were also fitted to the Si {\sc x} and S {\sc x} lines used to determine FIP bias.
This ratio shows some sensitivity to temperature and density, but our method is designed to remove these effects and the diagnostic has been proven to be robust in the composition analysis of a variety of features \citep[see references in the introduction and][ for a detailed discussion]{brooks11}.

The density was measured using the Fe {\sc xiii} line-pair diagnostic ratio.
The Fe {\sc xiii} ratio is highly sensitive to changes in density \citep{young07} and both lines are located in the 185--205 
$\angstrom$ wavelength range, where the EIS instrument is most sensitive.
These factors combine to make the Fe {\sc xiii} ratio the best EIS coronal density diagnostic \citep{young07}, however, its formation temperature is $\sim$0.4 MK higher than that of Si {\sc x} and S {\sc x} lines used in this study.
The difference in formation temperatures raises the possibility that Fe {\sc xiii} may be sampling different plasma to that of the Si {\sc x} and S {\sc x} emitting plasma. 
\citet{young09} conducted high precision density measurements using Fe {\sc xii} and Fe {\sc xiii} line pairs and found that the two different density diagnostics showed broadly the same trend in density across active regions. 
They concluded that any discrepancies in densities is most likely due to the atomic data for the ions rather than any real physical differences in the emitting plasmas. 
Subsequently, Fe {\sc xii} atomic data has been updated and improved by \citet{delzanna12} so that now the densities are in agreement for the two ions (Fe{\sc xii} and Fe {\sc xiii}).
These results provide solid evidence that the \emph{Hinode}/EIS is likely to be sampling the same plasmas since Si {\sc x}, S {\sc x}, and Fe {\sc xii} have the same formation temperature of 1.6 MK.

The CHIANTI Atomic Database, Version 8.0 \citep{dere97,delzanna15} was then used to carry out the calculations of the contribution functions, adopting photospheric abundances of \cite{grevesse07} for all of the spectral lines while assuming the previously measured densities.
The emission measure distributions were computed from the Fe lines using the Markov-Chain Monte Carlo (MCMC) algorithm available in the PINTofALE software package \citep{kashyap00} and then convolved with the contribution functions and fitted to the observed intensities of the spectral lines from the low-FIP element Fe.
As Si is also a low-FIP element, the emission measure was scaled to reproduce the Si {\sc x} line intensity.
FIP bias was then the ratio of the predicted to observed intensity for the high-FIP S {\sc x} line.
If the emission measure scaling factor of the Si {\sc x} line intensity derived from the Monte Carlo simulations to fit the observed line intensity was larger than the approximately 22$\%$ intensity calibration uncertainty then the pixels were excluded from consideration in our study \emph{e.g.} R3 in Figure \ref{fig:obs}(h) and R6 in Figure \ref{fig:obs}(f) \citep{lang06}.  
The scaling factor helps to account for any absolute calibration errors between the long wave and short wave detectors of \emph{Hinode}/EIS. 
A full account of the method is available in \cite{brooks15}.

To determine the mean FIP bias values within EFR R1--R7, profiles for each spectral line were averaged across all of the EFR pixels that were identified by the histogram-based intensity threshold technique in Figure \ref{fig:mask}.  
Although spatial information was lost when the profiles were averaged, the signal to noise was enhanced.
This is a necessary trade-off when measuring intensities within low emission regions such as EFRs in CHs.
The averaged profiles were fitted with single or multiple Gaussian functions and then the steps described above were carried out.
For the spatially resolved composition map in Section \ref{results2}, the method was also applied to each pixel within the map; therefore, no averaging was performed.
In this way, we retained the details of the FIP bias distributions within the larger EFRs (R1, R2).

Uncertainties in the FIP bias measurements are difficult to quantify since errors in the atomic data and radiometric calibration are likely to be systematic in nature. 
We conducted a series of experiments where intensities for a sample spectrum were randomly perturbed within the calibration error, and the standard deviation from a distribution of 1000 Monte Carlo simulations was calculated for a single pixel.
These experiments produced an uncertainty of $\sim$0.3 in the absolute FIP bias for single pixels within the composition map.

When the intensities of N pixels are summed the statistical error is expected to be reduced by a factor of 1/$\sqrt{N}$ for random fluctuations.
The mean FIP bias values of EFR R1--R7 were calculated over a range of N = [50, 2154], therefore, the corresponding reduced uncertainties were in the range [0.04, 0.006] in mean FIP bias.
Moreover, we investigated how FIP bias varied from one measurement to the next using a series of sit $\&$ stare observations of an active region where \emph{Hinode}/EIS is operating in the fixed target mode rather than rastering mode.
These observations provide a measure of how FIP bias changed with time over a period of $\sim$3.7 hours (each exposure lasted for 60 seconds and the total number of exposures was 220). 
The mean variation per pixel was 0.07.
These tests show that the uncertainties due to random fluctuations and variations from exposure to exposure are probably below to 0.1 for the relative abundance measurement per pixel.
For the mean FIP bias of an EFR, following the above argument, the statistical error is expected to be much lower than 0.1.
We still use this conservative error estimation in Figure \ref{fig:plots} since FIP bias measurements are highly complex.

\subsection{Magnetic Field Measurements}
A series of SDO/HMI line-of-sight magnetograms was used to determine emergence times, peak magnetic flux, magnetic flux density and estimates of loop length for the bipoles. 
The studied time period was from 00:00 UT on 2 January 2016 to 00:00 UT on 8 January 2016 using magnetograms at a cadence of 30 minutes.

An automated procedure, adapted from the method of \cite{james17}, was used to measure the magnetic flux associated with the regions of the emerging bipoles. 
Contours were drawn at $\pm$30 G on each line-of-sight magnetogram so that only pixels with magnetic flux densities greater than 30 G in magnitude were considered, and a further area threshold of 10 pixels was set so that only locations within contours that contained 10 or more pixels were used. 
These criteria combined to exclude weak field as well as small patches of pixels dominated by noise. 
The total magnetic flux within the contours that satisfied both threshold criteria was then calculated for each magnetogram for each bipole region.
Only the negative polarity was used for measurements of magnetic flux because the bipoles emerged in a positive-polarity CH, making it difficult to separate the positive flux associated with the emerging bipoles from the background CH field. 

\begin{figure}
	\centering
    \includegraphics[width=0.5\columnwidth]{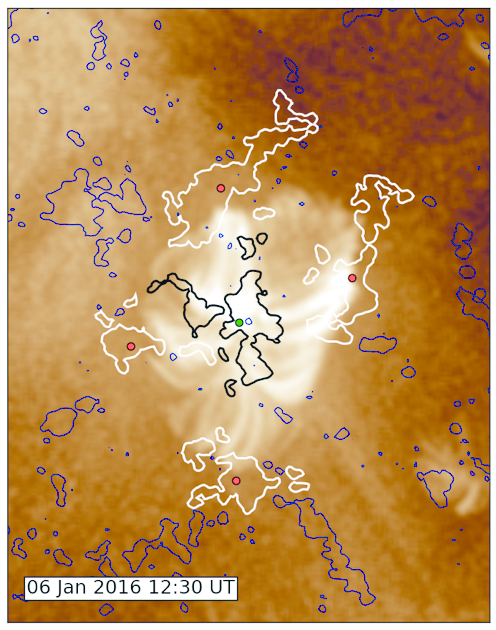}
    \caption{SDO/AIA 193 $\angstrom$ high resolution image overlaid with SDO/AIA magnetic field contours (positive/negative $\pm$30 G) of EFR R3.  Four groups of loops connecting the negative polarity with four separate positive polarities are identified.  Flux-weighted centroids of both polarities are indicated by the dots (green/red are negative/positive centroids). Loop lengths were then determined for each set of loops using the distance between the centroids.}
    \label{fig:loops}
\end{figure}

Values of magnetic flux density and estimated bipole loop length were determined at the times that FIP measurements were made. 
Loop length estimates were made by taking the flux-weighted centres of the positive and negative parts of the bipoles, and then using the separation of the centroids to calculate the lengths of semi-circular loops. 
For EFRs R3--R5 and R7, SDO/AIA 193 $\angstrom$ images overlaid with SDO/HMI magnetic field contours were used to identify different loop connectivities within the EFRs (see an example of R3 in Figure \ref{fig:loops}).
Distances between the negative and positive centroids for each set of loops were determined.
Because the positive flux had to be considered for the loop length estimates, the bipoles were isolated from any background CH field. 
This was done for both polarities using the magnetograms closest to the times of the FIP bias measurements.  
Only the contours that were determined by eye to enclose flux that corresponds to the desired bipole emergence were selected. 
Magnetic flux densities were also calculated for the negative polarity using these selected contours near the FIP measurement times by summing the magnetic flux density values within the selected contour and dividing by the total area within the contour.


\begin{table}
	\centering
	\caption{
Lifetime, mean FIP bias measurements for Regions R1--R7 (ranked by peak flux), time from emergence in hours, peak negative flux values for each flux region, and phase of flux region's life cycle. 
In the first column, estimated lifetimes of R4--R7 are provided in brackets after the region numbers.  R1--R3 rotated over the limb after their peak flux but before they dispersed into the background field, therefore their lifetimes cannot be estimated.
 Estimated uncertainty in mean FIP bias is $<$0.1 (see Section \ref{fipmeasure}).
 Time from emergence was calculated from when the regions emerged in HMI magnetograms until they were observed in \emph{Hinode}/EIS rasters.   
 The peak flux of R1 is given as of 00:00 UT on 8 January but the true peak occurred after the EFR rotated over the West limb.
 Emergence/decay phase is before/after peak flux. (E), (M), and (L) refer to early, middle, and late, respectively, of emerging or decay phases. 
    }
	\label{tab:results}
	\begin{tabular}{lcccr} 
		\hline
		Region       & FIP& Hours from & Peak Flux & Phase\\
          (lifetime) &Bias & Emergence& ($10^{19}$ Mx) & \\
		\hline
1 (NA)& 1.8 & 16.5 &      & Emergence ~(E)\\
         & 1.8 & 18.7 &      & Emergence ~(E)\\
         & 1.9 & 20.8 &  <-380 & Emergence ~(E)\\
        \hline
2 (NA)& 1.5 & 23.0 &      & Emergence (M)\\
         & 1.7 & 25.2 &      & Emergence (M)\\
         & 1.7 & 27.3 &  -16 & Emergence (M)\\
	     & 1.8 & 49.7 &      & Decay (M)\\
         & 1.8 & 51.8 &      & Decay (M)\\       
		\hline
3 (NA)& 1.8 & 18.5 &  -15 & Decay (E)\\
		 & 1.9 & 20.7 &      & Decay (E)\\
         & 1.8 & 22.8 &      & Decay (E)\\
         & 1.7 & 43.0 &      & Decay (L)\\
         & 1.2 & 47.3 &      & Decay (L)\\      
		\hline
4 (83 hrs)& 2.0 & 52.8 & -8.4 & Decay (L)\\
         & 1.8 & 54.9 &      & Decay (L)\\
         & 1.9 & 57.0 &      & Decay (L)\\
 	     & 1.9 & 77.0 &      & Decay (L)\\
		 & 1.7 & 79.2 &      & Decay (L)\\
         & 1.5 & 81.3 &      & Decay (L)\\    
		\hline
5 (31 hrs)& 1.7 & 16.3 & -3.1 & Decay (E)\\
         & 2.1 & 18.4 &      & Decay (E)\\
         & 2.5 & 20.5 &      & Decay (E)\\      
		\hline
6 (33 hrs)& 1.2 &  3.5 & -2.1 & Emergence (L)\\       
		\hline
7 (22 hrs)& 1.7 & 11.3 & -1.3 & Emergence (L)\\
         & 2.0 & 33.7 &      & Decay (L)\\
         & 1.6 & 35.8 &      & Decay (L)\\      
        \hline
	\end{tabular}
  
\end{table}


\section{Results}\label{results}
\subsection{Evolution of mean FIP bias in Emerging Flux Regions R1--R7}\label{results1} 
\begin{figure}
	\centering
    \includegraphics[width=0.5\columnwidth]{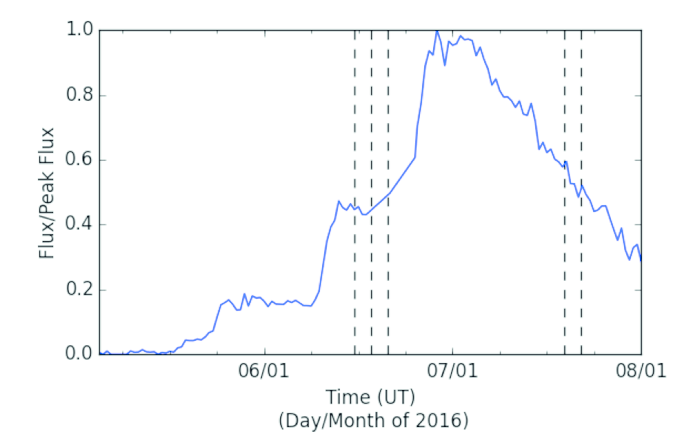}
    \caption{Plot of the total negative magnetic flux normalized to peak flux vs. time.  Dashed vertical lines indicate times of FIP measurements for R2  (See Table \ref{tab:results}).  The comparable length of the emergence and decay phases is typical of smaller EFRs.  }
    \label{fig:r2_bfield}
\end{figure}

Our results are tabulated in Table \ref{tab:results} where we show the mean FIP bias, hours from emergence in HMI magnetograms (called `age'), peak flux, and emergence/decay phase for EFRs R1--R7. 
Peak negative flux ranged from 1.3 $\times$ 10$^{19}$ (R7) to 3.8 $\times$ 10$^{21}$ Mx (R1).
The EFRs are classified as ephemeral regions, R4--R7, small active regions without spots, R2--R3, and an active region with spots designated AR 12481, R1  \citep[\emph{cf. }][]{lvdg15}.
With the exception of R1, the largest region, all EFRs had short lifetimes from one to four days and emergence/rise phases lasting for 0.25 to 3.5 days. 
R1 emerged two days before it rotated over the western limb, therefore, we were unable to determine its peak flux or lifetime.
The normalized flux profile of R2 in Figure \ref{fig:r2_bfield} shows some of the typical characteristics of flux evolution in small regions.
The emergence/decay phases are of nearly comparable duration, an important difference compared to ARs where the decay phase is much longer than the emergence phase.


The lowest mean FIP bias of 1.2 (\emph{i.e.,} nearly photospheric composition) occurred in R6, 3.5 hours after it emerged in the CH and in R3 during the late decay phase, two days after emergence.
EFR R5 had the highest value of 2.5, which is indicative of enriched plasma (compared to quiet Sun FIP of $\sim$ 1.5), and it was measured during the decay phase $\sim$20.5 hours after emergence.
We observed three other EFRs (R1, R2, and R7) during their emergence phases and all had enhanced mean FIP bias, $>$1.5 at 11--27 hours after emergence, assuming photospheric composition (FIP bias of 1) at the time of emergence.
In general, mean FIP bias was enhanced to levels greater than that of the quiet Sun (FIP bias of $\sim$1.5) less than a day from the beginning of flux emergence in HMI magnetograms.  

\begin{figure*}
	\centering
    \includegraphics[width=4.6in]{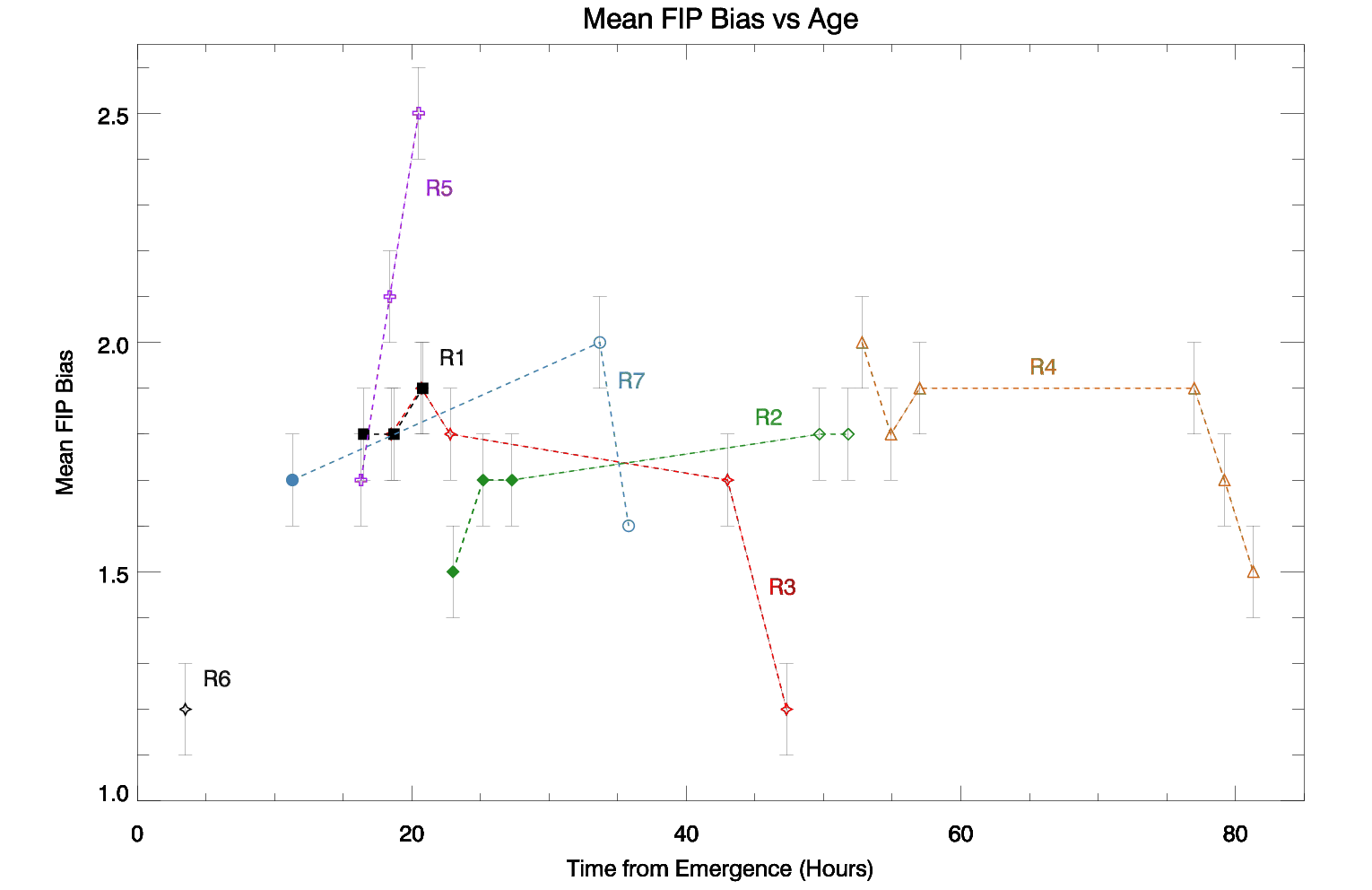}
    \includegraphics[width=4.6in]{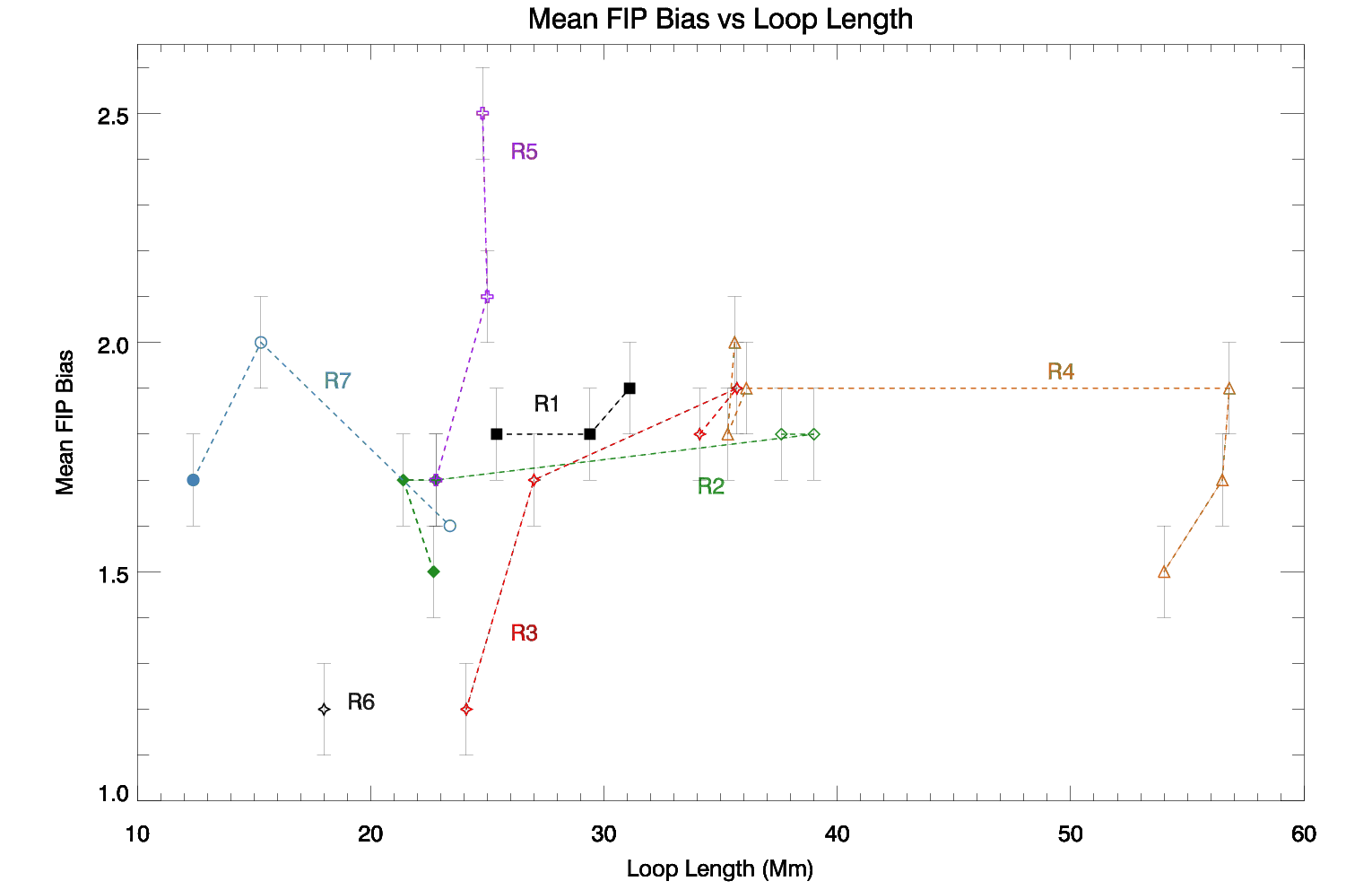}
    \includegraphics[width=4.6in]{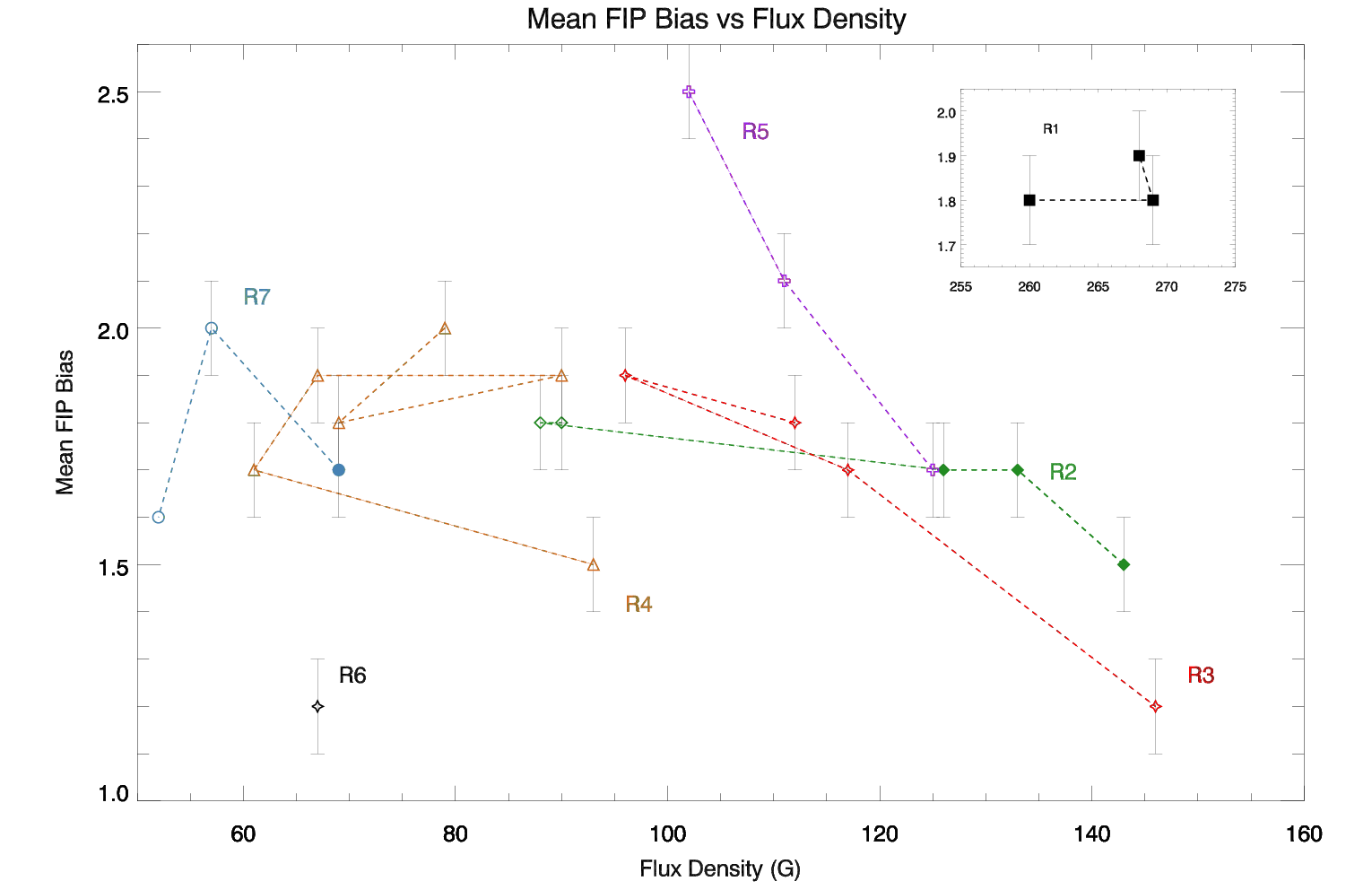}
  \caption{(a)  Mean FIP bias vs. time from emergence (\emph{i.e.,} age), (b)  mean FIP bias vs. loop length (Mm), (c)  mean FIP bias vs. magnetic flux density.  Filled/open symbols represent times of mean FIP bias measurements prior to/after peak magnetic flux.  R1--R7 refer to regions labeled in Figure \ref{fig:obs}.  Inset in (c) shows R1 separately to improve the display of the other regions. Uncertainty is $<$ 0.1 in mean FIP bias (see Section \ref{fipmeasure}).
  }
    \label{fig:plots}
\end{figure*}

\emph{Hinode}/EIS observed the evolution of mean FIP bias in six of the seven EFRs for periods lasting from a few hours to two days.
Figure \ref{fig:plots} (a) shows mean FIP bias vs. time from emergence (age).  
The values in the plots are from Table \ref{tab:results} and were determined using the average of N pixels within each region.  Therefore, the uncertainty is estimated to be $<$0.1 in mean FIP bias (see Section \ref{fipmeasure}). 
Filled/unfilled symbols denote the emergence/decay phases of the EFRs (\emph{cf.} Table \ref{tab:results}).
Three EFRs (R3, R4, and R7) had similar patterns of mean FIP bias evolution during the EIS observing period.
They maintained steady plasma composition for approximately one day before declining to levels that are more typical of either quiet Sun ($\sim$1.5 for R4 and R7), the so-called basal state \citep{ko16}, or photospheric FIP bias levels of the surrounding coronal hole (R3).
Next, EFR R2 had increasing mean FIP bias for a day, however, we are unable to comment on later evolution because \emph{Hinode}/EIS no longer observed the CH after the last raster at 15:15 UT on 7 January.

EFR R5 exhibited anomalous mean FIP bias evolution in comparison to the other regions.
In the first observation, R5 appeared to have reached an enriched plasma composition on a similar time scale as all other regions, however, R4 was not observed early enough to draw any conclusions.
Mean FIP bias then rapidly increased from 1.7 to 2.5 in 4.2 hours to reach the highest level in our study.
The plasma is enhanced at a rate of 0.2 h$^{-1}$ in mean FIP bias. 
In contrast, mean FIP bias increased by an order of magnitude slower in the range of [0.03, 0.06] h$^{-1}$ for R1--R3, and R7 (based on the change from photospheric FIP bias of 1 at the first magnetic appearance up to the emergence/early decay phase of each region).
Finally, mean FIP bias decreased by $\sim$0.1--0.2 h$^{-1}$ in R3, R4, and R7 (based on the sharp decrease in mean FIP bias for the last 2--3 measurements).

There was no real change in the mean coronal composition of R1 since it was observed for a short time interval.  
Mean FIP bias was 1.8--1.9 for 4.3 hours on 7 January.
This region was the largest in our sample and was in the very beginning of its emergence phase, unlike the other regions which were observed from the late stages of emergence to late decay phases.

We investigated mean FIP bias parameter space and found no apparent trends in mean FIP bias vs. peak magnetic flux, normalized peak magnetic flux or  emergence time \emph{i.e.,} the time period from the beginning of emergence to peak flux.
Next, we tested the possible dependence of the mean FIP bias on two parameters, loop length and magnetic flux density.
Figure \ref{fig:plots} (b) shows mean FIP bias vs. loop length (Mm). 
For EFRs with multiple loop connectivities, the plotted loop length is an average value of loops within the EFR at the time of the mean FIP bias measurements.    
The number of loop groups per region ranged from 2--5 with variances in the loop length as small as $\pm$5 Mm for R4 and as large as $\pm$19 Mm for R5.
In our sample and within the estimated uncertainties, there is no FIP bias dependence with loop length.
Mean FIP bias vs. magnetic flux density is displayed in Figure \ref{fig:plots} (c).
Overall, the graph shows a decreasing trend in mean FIP bias with increasing flux density (with the exception of R1, see inset in Figure \ref{fig:plots} (c), where values are too close to derive a reliable trend).

\begin{figure*}
	\centering
 \includegraphics[width=7in]{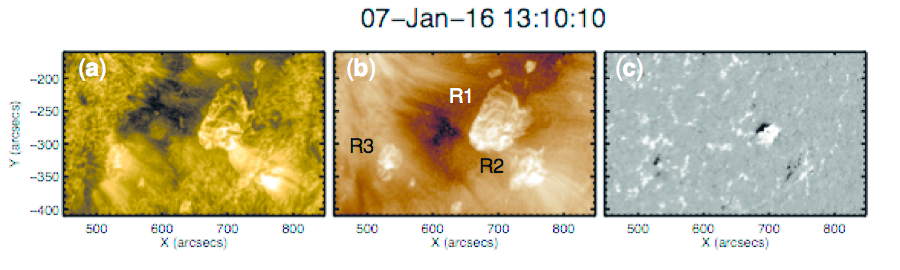}
    \caption{SDO/AIA 171 $\angstrom$ (a) and 193 $\angstrom$ (b) images, and SDO/HMI magnetogram (c) of EFRs R1, R2, R3 at 13:10 UT on 7 January.  Images are taken from Movie 2.   
    }
    \label{fig:movie2}
\end{figure*}

\begin{figure*}
	\centering
    \includegraphics[width=7in]{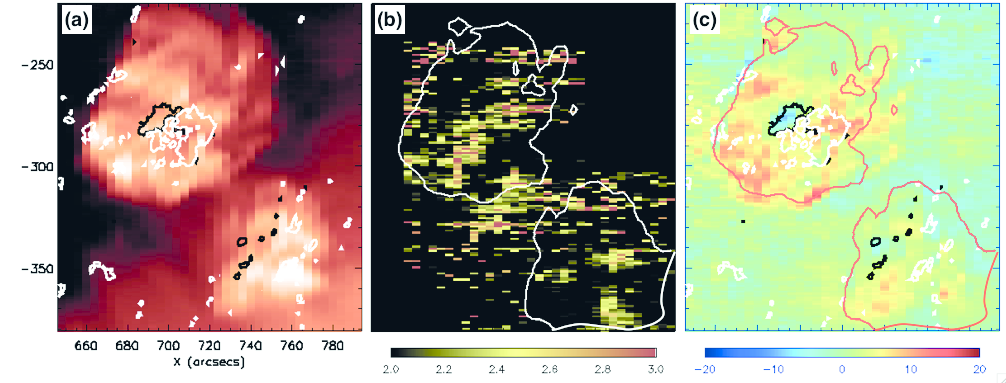}
    \caption{\emph{Hinode}/EIS Fe {\sc xii} 195~\angstrom~intensity map overlaid with $\pm$100 G magnetic flux contours (a), composition map (b) showing only FIP bias values greater than 2 overlaid with contours of the masks determined using the intensity threshold method (see Section \ref{region_define}), and Fe {\sc xii} 195~\angstrom~Doppler velocity map (c) overlaid with magnetic flux and mask contours.  Regions 1 and 2 at 13:09 on 7 January.  Estimated uncertainties within the composition map is $\pm$0.3 per pixel (see Section \ref{fipmeasure}).
    }
    \label{fig:mini_fip2}
\end{figure*}

\subsection{Plasma Composition Map}  \label{results2}
Typically, emergence of a bipole within a CH leads to the formation of an `anemone' structure in the corona;
see Figures \ref{fig:movie2} (a) and (b). 
The emerging flux interacts with the surrounding CH field via interchange reconnection of oppositely directed field. 
New loops extend radially from the location of the included EFR polarity. 
Movie 2 shows the three largest regions R1, R2 and R3 evolving into classic anemone regions in SDO/AIA 171~$\angstrom$~and 193~$\angstrom$~passbands.

EFR R1 emerged in the open magnetic field environment of the CH 33 hours after EFR 2 which emerged at the boundary between the CH and quiet Sun.
The natural expansion of the growing EFRs leads to new, extended loops forming at the interface of the two regions where the magnetic field was oppositely aligned. 
Movie 2 and Figure \ref{fig:movie2} (a,b) show the coronal interconnectivity of the EFRs. 
Figure \ref{fig:mini_fip2} displays \emph{Hinode}/EIS zoomed Fe {\sc xii} intensity and Doppler velocity maps and Si {\sc x}--S {\sc x} composition map for R1 and R2 on 7 January (\emph{cf.} Figure \ref{fig:obs} (h)).
The CH appears dark in the intensity map whereas the EFRs are bright, compact features. 
The corresponding locations in the Doppler velocity map show plasma flows almost at rest with isolated patches of blue-shifted plasma flows of 10--20 km s$^{-1}$ along the open field of the CH, and red-shifted plasma flows of 10--20 km s$^{-1}$ contained within the closed loops of the EFRs. 
These are typical velocity ranges for CHs and quiescent ARs.

The composition map show FIP bias values greater than 2 so that locations of enriched FIP bias are clearly visible within the two EFRs.
Coronal hole and quiet Sun pixels have FIP bias $<$2. 
Contours of the masks described in Section \ref{region_define} are overlaid on the composition and Doppler velocity maps.
Within the contours, some of the bright loops contain plasma enriched well above quiet Sun FIP bias of $\sim$1.5.
Moreover, where the EFRs connect with each other (R1 and R2), FIP bias is once again enriched.

\begin{figure*}
	\centering
    \includegraphics[width=7in]{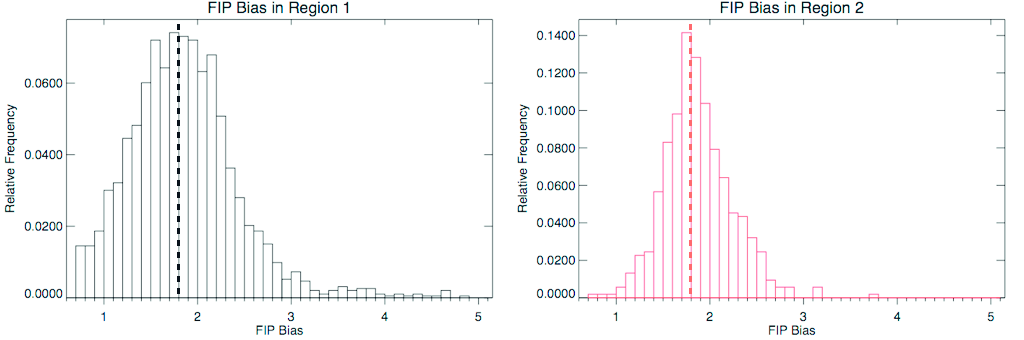}
    \caption{Histograms of FIP bias in regions R1 and R2 defined with intensity thresholds and shown in composition map in Figure \ref{fig:mini_fip2}.  Dashed lines correspond to mean FIP bias values. 
    }
    \label{fig:hist}
\end{figure*}

Figure \ref{fig:hist} shows the relative frequency distributions of FIP bias within the mask contours for EFRs R1 and R2.
The distributions are skewed to higher FIP bias values and the percentages of pixels of FIP bias greater than 2 are 36$\%$ and 32$\%$, respectively.  
The spatially resolved composition map and the frequency distributions clearly show that FIP bias varies from photospheric up to coronal values across EFRs of different sizes.  

\section{Discussion}  \label{disc}

\subsection{Comparison of EFRs and ARs} 
The EFRs in our study are at the small end of the size-spectrum of ARs \citep{lvdg15}.
Only the stronger magnetic fields of the largest EFRs, R1--R3, are organized according to Hale's polarity law.
The smaller regions, R4--R7, have anti-Hale orientation with a large scatter of inclination angles as they are presumably too small to resist the buffeting effects of convection \citep{longcope96}. 

The small magnetic flux content of EFRs has implications for the ratio of the length of the emergence phase to the lifetime of the AR.
\cite{harvey93} and \cite{lvdg15} show that for EFRs comparable to the ones in our study, the duration of the emergence phase is $\sim$30$\%$ of the lifetime. 
For full-sized ARs, the percentage goes down to $<$15$\%$.
As AR size decreases, the asymmetry of the emergence-to-decay phases also decreases so that the ephemeral-like EFRs are close to having emergence and decay phases of comparable time periods (\emph{cf.} Figure \ref{fig:r2_bfield}).

\subsection{Emergence/early decay phase} 
\emph{Skylab} spectroheliogram observations showed that the average composition of ARs is photospheric just after emergence \citep[\emph{e.g., }][]{sheeley95,sheeley96}.
There is a progressive enrichment of the plasma at an almost constant rate in the evolving regions so that within two days, the plasma is modified to mean FIP bias $\sim$3--4 and after a week, it exceeds 7 \citep{widing01}.
In the observations shown here, plasma composition in five of seven EFRs within the CH had evolved to levels exceeding quiet Sun values ($>$1.5) in less than a day from emergence (see Table \ref{tab:results} and Figure \ref{fig:plots}).

Based on the length of the emergence phases of the four ARs in the \emph{Skylab} study (2--5 days), we estimate that their maximum magnetic flux range from $5 \times 10^{21}$ to $10^{22}$ Mx \citep{lvdg15}.
They would be classified as large ARs with lifetimes of weeks to months.
Here, the EFRs were much smaller by as much as 1--2 orders of magnitude with the exception of R1 ($3.8 \times 10^{21}$~Mx), so that their lifetimes were measured in days.
 
In the four ARs observed by \emph{Skylab}, the mean FIP bias ratio increases at a rate of approximately [0.04, 0.1]~h$^{-1}$ and is maintained for 5--7 days during their emergence phases.
Four flux regions in our study exhibited comparable positive rates of enrichment [0.03, 0.06]~h$^{-1}$ during their emergence/early decay phases.
The rates of enrichment determined from our observations are based on the assumption that the EFRs emerged with photospheric plasma (FIP bias =1).
Though no EFR was captured by \emph{Hinode}/EIS precisely at the time of emergence, this is a reasonable assumption based on the early \emph{Skylab} results \citep{sheeley95,sheeley96,widing97,widing01}. 
In our sample, FIP bias was 1.2 in R6 within 3.5 hours from emergence which is consistent with \emph{Skylab} results.

Any discussion of FIP bias evolution is complicated by the behavior of the element S depending on factors such as the height in the chromosphere at which fractionations occurs.
The Laming model \citep{laming12} predicts that S acts as a high-FIP element at the top of the chromosphere, whereas it is more fractionated, behaving like a low-FIP element, at lower heights.
Observationally, \citet{lanzafame02} showed that S behaves like a high-FIP element in active regions whereas \citet{brooks09} demonstrated that it is more like an 'intermediate' FIP element in the quiet-Sun.
See \citet{baker13} for an extended discussion.

The evolution of plasma composition from emergence in the EFRs is likely to be far more complex than the simple linear trend assumed here.
These are short-lived, relatively small flux regions which are located in a complex environment within or nearby a coronal hole.
Further \emph{Hinode}/EIS observations are needed to test FIP bias values at the beginning of the emergence phase. 

\subsection{Late decay phase} 
For those EFRs where we have sufficient measurements over time, mean FIP bias decreased during the late decay phase for a further 1--2 days while the magnetic field of the EFRs was dispersing.
The processes in the decay phase involve the interaction between supergranular cells and the dispersing magnetic flux. 
The convection-buffeted flux concentrations become confined to inter-granular lanes, outlining the ever-evolving supergranular cells. 
Along the supergranular boundary lanes opposite-polarity flux concentrations meet and cancel, a process which effectively and quickly removes the minority-polarity flux of these EFRs (\emph{i.e.,} negative polarity in this CH). 

In a large AR, \cite{baker15} demonstrate that AR plasma with mean FIP bias $\sim$3 may be modulated by small bipoles emerging at the periphery of supergranular cells containing photospheric material reconnecting with pre-existing AR field.
A similar scenario is likely to affect the observed mean FIP bias levels in the smaller EFRs as the decay processes are not unique to large ARs, in particular interaction and plasma mixing with surrounding fields.
Furthermore, \cite{ko16} show a temporal evolution from moderate to strong positive correlation of FIP bias and the weakening photospheric magnetic field strength during the decay phase of a large AR, with the largest  decrease in FIP bias occurring in plasma at $\sim$2~MK.
Once the photospheric magnetic field evolves to below 35 G, FIP bias in the AR reaches a quiet Sun basal state for FIP bias $\sim$1.5.  The correlation of FIP bias and magnetic field strength in their study does not hold below 10 G which is typical mean field strength for CHs.

In our study, the rates of change in mean FIP bias in the small flux regions during their late decay phase are significantly faster than the rates of change observed in larger decaying ARs.
\cite{baker15} and \cite{ko16} find remarkably similar rates of decreasing mean FIP bias in the range [0.004, 0.009] h$^{-1}$ within the core of ARs over 2--5 days compared to a range of [0.1, 0.2] h$^{-1}$ for R3, R4, and R7 during their late decay phase.
Once again, the rates of decrease exhibited in the large ARs are maintained over periods of days, whereas, in the small regions within the CH, the rapid changes occurred in a matter of hours.

\subsection{Spatial distribution of FIP bias} 
The spatially resolved composition map in Figure \ref{fig:mini_fip2} as well as in previous studies \citep[\emph{e.g., }][]{baker13,baker15,brooks15}, have revealed that FIP bias does vary over spatial scales of a few arc seconds.
Patches of higher FIP bias ($>$2) were found in the core of the EFRs which is consistent with the \emph{Hinode}/EIS observations of large ARs in \cite{baker13,baker15}. 

There was enriched plasma in the vicinity of loops linking the positive polarity of R1 and the negative polarity of R2  (see patches of FIP bias in the range [2, 3] between the mask contours of Figure  \ref{fig:mini_fip2} (b) which spatially correspond to the loops connecting R1 and R2 in Figure \ref{fig:movie2} (a,b) and Movie 2).
The magnetic field alignment of R1 relative to R2 was favourable for reconnection between external loops of each region. 
The reconnecting loops may contain already enriched plasma but it is also possible that the Alfv\'en waves generated by reconnection may have stimulated the enrichment \citep[\emph{e.g., }][]{laming04,laming15}.

Finally, plasma composition within the two largest EFRs had non-Gaussian distributions ranging from photospheric to coronal FIP bias (Figure \ref{fig:hist}).  
The distributions were mildly skewed towards the higher end of the distributions with 32--36$\%$ of the pixels within the masked regions exceeding FIP bias = 2. 
The lower end of the range of FIP bias values was consistent with previous observations of FIP bias in newly emerged loops \citep{sheeley95,sheeley96,widing01,laming15} and with studies of plasma composition in CHs \citep{feldman98,brooks11,laming15}, whereas the upper end of the range, $\sim$3 was lower than the levels typically found in quiescent AR cores of $\sim$3--4 \citep[\emph{e.g., }][]{baker15,delzanna15}.
The fraction of pixels containing FIP bias $>$3 was small in these regions R1: $\sim$4$\%$ and R2: $\sim$1$\%$.
The extent to which plasma is enriched is likely to be affected by the CH environment.
EFRs are partially or fully surrounded by CH field containing a large reservoir of unmodified photospheric-composition plasma.
Interchange reconnection between the closed loops of the EFRs and the open field of the CHs creates pathways for mixing of coronal and photospheric plasmas, and as a consequence, the enhancement of FIP bias may be modulated \citep{baker13,baker15}. 

\section{Conclusions}\label{conclusion}
We analyzed the temporal evolution and spatial distribution of coronal plasma composition within flux regions of varying sizes located inside an equatorial coronal hole using observations from \emph{Hinode}/EIS.
We obtained FIP bias measurements in seven emerging flux regions during different phases of their lifetimes.
In general, plasma was enriched from a photospheric level to values greater than the quiet Sun in less than one day from initial flux emergence.
FIP bias remained steady for 1--2 days before declining during the decay phase to the photospheric composition of the surrounding coronal hole.

The spatially resolved composition map revealed how FIP bias was distributed in and around the small flux regions located in the CH.
Plasma containing FIP bias in the range 2--3$^{+}$ was concentrated in core loops and in the area where interconnecting loops were formed by reconnection between loops of neighboring EFRs.
At the interface between open and closed field, plasma mixing appeared to occur where reconnection took place between EFR closed loops containing enriched material and the surrounding coronal hole filled with photospheric plasma. 
We conclude that the variation in plasma composition observed in all sizes of flux regions from ephemeral to active regions is affected by the magnetic topology of each region in its surrounding environment.

The spatial distribution of FIP bias in the composition map of ephemeral-like EFRs is similar to that of the anemone AR in \cite{baker15} which is up to an order of magnitude larger in magnetic flux content.  
Conversely, the rate of composition change from coronal to photospheric observed during the magnetic decay phase of the EFRs is significantly faster compared to that of larger ARs \citep{baker15,ko16}.
Not only is the magnetic flux decay rate faster for the EFRs, but the so-called basal FIP bias level is photospheric in the coronal hole ($\sim$1), rather than $\sim$1.5 in the quiet Sun.
Furthermore, EFRs are readily reconnected with the surrounding field as their weak magnetic fields are more affected by the convection which disperses the magnetic field more efficiently than in ARs.  

In general, our results indicate that mean FIP bias increases during the magnetic emergence and early-decay phases, while it decreases during the magnetic late-decay phase.
The rate of increase during the emergence phase is likely to be linked to the fractionation mechanism and transport of fractionated plasma leading to the observed coronal FIP bias, holding true for three orders of magnitude in magnetic flux from ephemeral-like EFRs to large ARs.
On the other hand, the rate of decrease in mean FIP bias during the decay phase depends on the rate of reconnection with the surrounding magnetic field and the composition of the surrounding corona. 

\acknowledgments
The authors would like to thank Dr. Martin Laming for his insightful discussions of our observations.
{\it Hinode} is a Japanese mission developed and launched by ISAS/JAXA, collaborating with NAOJ as a domestic partner, NASA and STFC (UK) as international partners. 
Scientific operation of {\it Hinode} is by the {\it Hinode} science team organized at ISAS/JAXA. 
This team mainly consists of scientists from institutes in the partner countries. 
Support for the post-launch operation is provided by JAXA and NAOJ (Japan), STFC (UK), NASA, ESA, and NSC (Norway). 
SDO data are courtesy of NASA/SDO and the AIA and HMI science teams.
DB is funded under STFC consolidated grant number ST/N000722/1.
The work of DHB was performed under contract to the Naval Research Laboratory and was funded by the NASA Hinode program.
LvDG is partially funded under STFC consolidated grant number ST/N000722/1. 
LvDG also acknowledges the Hungarian Research grant OTKA K-109276.
AWJ acknowledges the support of the Leverhulme Trust Research Project Grant 2014-051.
DML is an Early-career Fellow funded by The Leverhulme Trust.


\bibliographystyle{aasjournal}

\end{document}